\documentclass[12pt]{iopart}

\usepackage{float}
\usepackage[english]{babel}
\usepackage{upgreek}
\usepackage{dsfont}
\usepackage{cite}
\usepackage{graphicx}
\usepackage[dvipsnames]{xcolor}

\usepackage{bm}

\newcommand{\op}[1]{{\sf #1}}

\newcommand{\cL}{{\mathcal L}}

\newcommand{\oH}{{\mathsf H}}

\newcommand{\eps}{\varepsilon}

\newcommand{\ket}[1]{\left \vert #1 \right \rangle}
\newcommand{\braket}[2]{\langle #1 \vert #2 \rangle}
\newcommand{\matel}[3]{{\left\langle \vphantom{#1 #2 #3} #1 \,\right\vert
\left.
 \hspace{-0.15em} \vphantom{#1 #2 #3} #2 \,\right\vert \left.
 \hspace{-0.15em} \vphantom{#1 #2 #3} #3\right\rangle}}
\newcommand{\ketbra}[2]{\vert #1 \rangle \langle #2 \vert}
\newcommand{\be}{\begin{equation}}
\newcommand{\ee}{\end{equation}}
\newcommand{\bea}{\begin{eqnarray}}
\newcommand{\eea}{\end{eqnarray}}

\newcommand{\mitl}[1]{\left \langle #1 \right \rangle}
\newcommand{\un}{{\mathds 1}}

\newcommand{\oJ}{\textsf{\textbf{J}}}

\newcommand{\oOmega}{\op{\Omega}}

\newcommand{\oV}{\op{V}}

\newcommand{\rR}{\mathrm{R}}

\renewcommand{\color}[1]{}  

\begin{document}
\title{Probing Macroscopic Quantum Superpositions with Nanorotors}

\author{Benjamin A. Stickler$^1$, Birthe Papendell$^1$, Stefan Kuhn$^2$, Bj\"orn Schrinski$^1$, James Millen$^{2,3}$, Markus Arndt$^2$, and Klaus Hornberger$^1$}

\address{$^1$ University of Duisburg-Essen, Faculty of Physics, Lotharstra\ss e 1, 47048 Duisburg, Germany}
\address{$^2$ University of Vienna, Faculty of Physics, VCQ, Boltzmanngasse 5, 1090, Vienna, Austria}
\address{$^3$ King's College London, Department of Physics, Strand, WC2R 2LS, London, UK}

\begin{abstract}
Whether quantum physics is universally valid is an open question with far-reaching implications. Intense research is therefore invested into testing the quantum superposition principle with ever heavier and more complex objects. Here we propose a radically new, experimentally viable route towards studies at the quantum-to-classical borderline by probing the orientational quantum revivals of a nanoscale rigid rotor. The proposed interference experiment testifies a macroscopic superposition of all possible orientations. It requires no diffraction grating, uses only a single levitated particle, and works with moderate motional temperatures under realistic environmental conditions. The first exploitation of quantum rotations of a massive object opens the door to new tests of quantum physics with submicron particles and to quantum gyroscopic torque sensors, holding the potential to improve state-of-the art devices by many orders of magnitude.
\\[\baselineskip]
{\bf Published in:\ } B. A. Stickler {\em et al.}, New Journal of Physics 20, 122001 (2018) 
\end{abstract}

\maketitle

\section{Introduction} 
Various experiments have been  carried out to test the validity of the quantum superposition principle \cite{arndt2014a} for ever heavier and more complex objects \cite{arndt1999,friedman2000,kohstall2011,eibenberger2013,berrada2013,kovachy2015,robens2015,riedinger2018,ockeloen2018}. Such experiments aim at probing modifications of linear quantum physics, have far-reaching technological applications, and may eventually reveal quantum aspects of gravity \cite{bose2017,marletto2017}. Most proposals for future experiments suggest observing the superposition of different motional states of micromechanical oscillators \cite{marshall2003,pikovski2012} or the center-of-mass interference of free-flying nanoparticles \cite{romeroisart2011a,scala2013,bateman2014,kaltenbaek2016,wan2016a,pino2018}. Here we propose an entirely new way of testing the quantum superposition principle by exploiting the quantized orientational degree-of-freedom of a rigid object.

The proposed interference scheme is based on the fact that an initially tightly oriented quantum rotor rapidly disperses, while  at multiples of a much longer quantum revival time the collective interference of all occupied angular momentum states leads to a complete re-appearance of the initial state \cite{seideman1999,poulsen2004}. Surprisingly, we find that such {\it orientational quantum revivals} can be probed with modest initial temperatures, involving hundreds of thousands of total angular momentum quanta, and under realistic environmental conditions, taking into account all relevant sources of orientational decoherence \cite{stickler2016b,zhong2016}. The proposed experiment enables the first test of macroscopic angular momentum quantization with massive objects. It requires no diffraction grating and uses only a single recyclable nanoparticle. The empirical measure of macroscopicity \cite{nimmrichter2013} (quantifying to which extent a superposition test falsifies a wide class of classicalizing modifications of quantum physics) is comparable to that of ambitious center-of-mass proposals.

We argue that macroscopic orientational quantum revivals can be observed using existing technology for optically manipulating the motion and alignment of levitated particles \cite{kane2010,kuhn2015,hoang2016,kuhn2017b}. The proposed scheme requires cavity- or feedback-cooling of the nanoparticle rotation \cite{stickler2016a,zhong2017} to below a Kelvin, while it is independent of its center-of-mass temperature. An observation of orientational quantum revivals with nanorotors would substantially advance macroscopic superposition tests, provide the first experimental test of the angular momentum quantization of massive objects, and enable quantum coherent gyroscopic torque sensing with the potential of improving state-of-the-art devices \cite{kuhn2017b,wu2017} by many orders of magnitude.

\section{Orientational Quantum Revival Scheme}

\begin{figure*}
  \centering
  \includegraphics[width=0.95\textwidth]{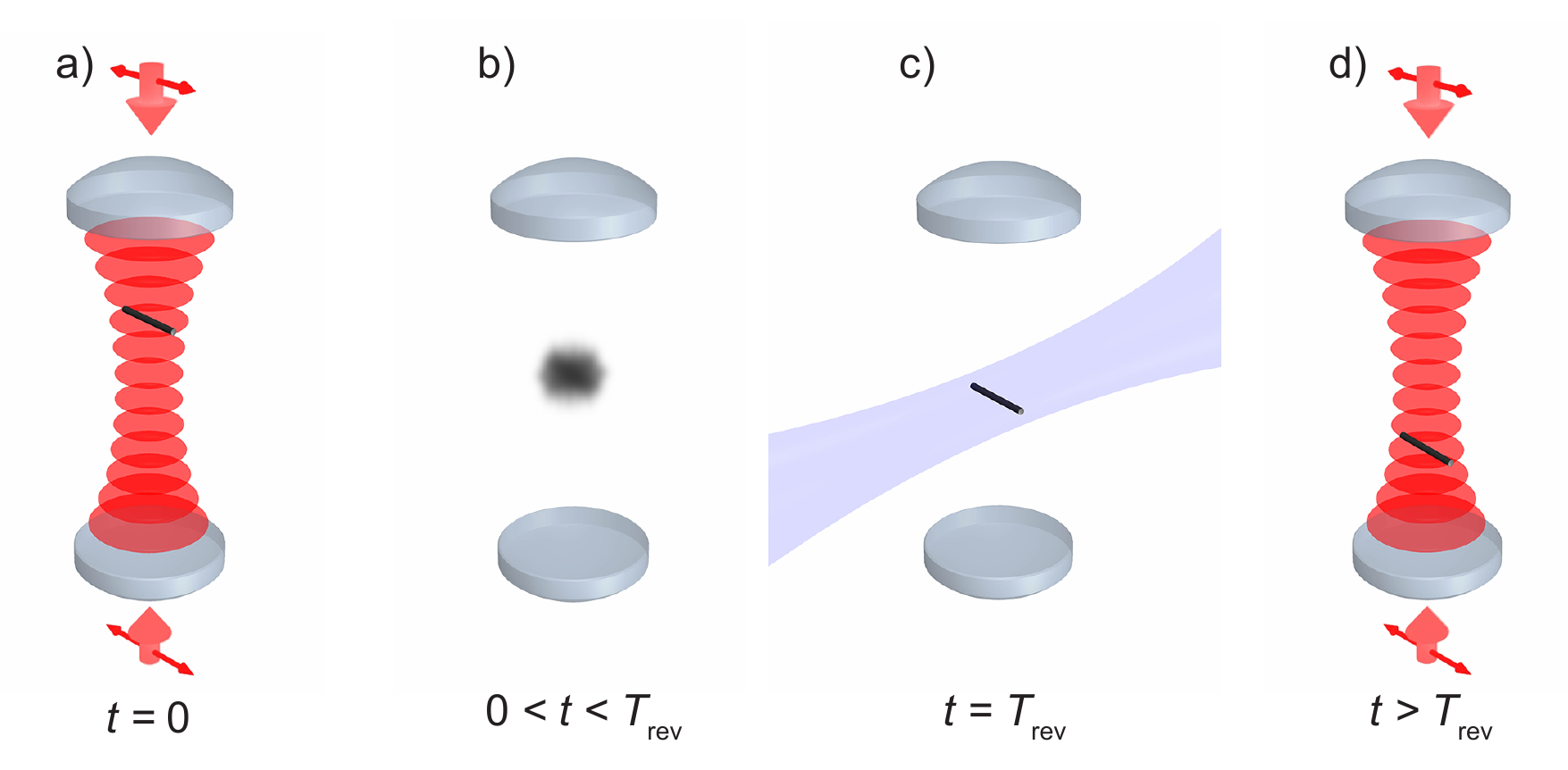}
  \caption{Scheme to observe orientational quantum revivals of a nanoscale rotor. (a) The rotor is levitated in an optical tweezer formed by two counter-propagating {\color{RoyalBlue}linearly polarized} laser beams. Cavity or feedback cooling to subkelvin temperatures tightly aligns it with the field polarization {\color{RoyalBlue}(indicated by the small arrows)}. (b) After switching off the trapping and cooling beam the quantum state of the rotor quickly disperses into a superposition of all possible orientations, while its center of mass drops in the gravitational field. (c) At integer multiples of the revival time $T_{\rm rev}$ the initial state is recovered by constructive interference  of all occupied rotation states. This \emph{orientational quantum revival} is detected by the total light scattered from a {\color{RoyalBlue}plane-wave} probe {\color{RoyalBlue}pulse}, which collapses the rotor into a state of definite orientation. (d) To repeat this quantum measurement several times, the rotor is recaptured by the trapping lasers and then transferred back to step (a) by tuning their relative phase.
}
\label{fig:fig1}
\end{figure*}

The proposed scheme consists of cycling the four consecutive steps displayed in Fig.~\ref{fig:fig1}: (a) alignment, (b) dispersion, (c) revival, and (d) recapture. To discuss each step in detail, we consider a nanoscale linear rigid rotor of length $\ell$, mass $M$, and moment of inertia $I = M \ell^2/12$ levitated in high vacuum by an optical tweezer, consisting of two counter-propagating beams of power $P$ and {\color{RoyalBlue}linear} polarization direction $\boldsymbol{\eps}$, which form a standing wave of waist $w$. The tweezer is aligned with the gravitational field so that the released particle drops along the tweezer axis. Denoting the angle between the rotor symmetry axis and $\boldsymbol{\eps}$ by $\beta$, the optical potential of the nanoparticle orientation at the antinode is given by $V(\beta) = - V_0 \cos^2 \beta$. The potential depth is $V_0 = 4 \Delta \alpha P/\pi c\epsilon_0 w^2 $ with the polarizability anisotropy $\Delta \alpha = \alpha_\| - \alpha_\bot$.

\subsection{Alignment} 

The orientation of the nanoparticle can be cavity or feedback cooled \cite{stickler2016a,zhong2017} leading to a tight alignment of the rotor with the field polarization direction, {\color{RoyalBlue}so that its trapped dynamics are librational rather than rotational}. The quantum state of the rotor is then characterized by its {\color{RoyalBlue}librational} temperature $T$ and takes the form $\rho_0 = \exp ( - \oH/k_{\rm B} T) / Z$ with the Hamiltonian $\oH = \oJ^2/2 I + V(\hat{\beta})$ and the partition function $Z$, where $\oJ$ is the angular momentum operator. (Operators are denoted by sans serif characters.)

The matrix elements of the initial state in the angular momentum basis $\ket{jm}$ (with $\hbar m$ the angular momentum component along the field polarization) can in principle be  evaluated numerically by exact diagonalization of $\oH$. However, in the semiclassical regime, where millions of angular momentum states are occupied, exact diagonalization becomes numerically intractable. Using Bohr-Sommerfeld-quantized action-angle variables \cite{child2014} the semiclassical matrix elements take the form (see \ref{app:A})
\begin{eqnarray} \label{eq:scis}\fl
\matel{j m}{\rho_0}{j'm'} \simeq & \frac{\delta_{mm'}}{Z} I_{({j-j'})/{2}}\left[\frac{V_0}{2k_{\rm B}T}\left(1-\frac{4m^2}{(j+j'+1)^2}\right)\right]
\mathrm{exp}\left[-\frac{\hbar^2 (j + j' + 1)^2 }{8Ik_{\rm B}T} \right ]
 \nonumber \\
& \times \exp \left [\frac{V_0}{2 k_{\rm B} T} \left ( 1 -  \frac{4 m^2}{(j + j'+1)^2} \right )\right]  
\end{eqnarray}
for $j - j'$ even and $\matel{j m }{\rho_0}{j'm'} = 0$ otherwise. Here, $I_n ( \cdot )$ denotes the modified Bessel function. The expectation value of the total angular momentum quantum number can be approximated as (see \ref{app:B}) $\mitl{j}_0 \simeq \sqrt{\pi k_{\rm B} T I/2 \hbar^2}$, yielding a mean occupation of $\mitl{j}_0 \simeq 2.6\times 10^4$ for $\ell = 50$~nm, $M = 10^6$~amu, and $T = 1$~K.

The initial orientational alignment can be quantified by the expectation value $\mitl{\cos^2 \beta}_0 = \mathrm{tr}( \rho_0 \cos^2 \hat{\beta})$, from now on referred to as the {\it alignment}. It is unity for a perfectly aligned particle and $1/3$ for uniformly distributed orientations. For the initial state Eq.~\eref{eq:scis} the alignment can be determined in leading order of $k_{\rm B} T/V_0$ as (see \ref{app:B})
\begin{equation} \label{eq:inalign}
 \mitl{\cos^2 \beta}_0 = k_{\rm B} T \frac{\partial}{\partial V_0} \ln Z  \simeq 1  - \frac{k_{\rm B} T}{V_0}.
\end{equation}
This relation holds if many angular momentum quanta are occupied (but fails in the deep quantum regime where the alignment is limited by the uncertainty relation).

\subsection{Rotation dynamics during free fall} 

Once the trapping laser is turned off, the orientation state evolves freely while its center of mass drops in the gravitational field along the tweezer axis. The ensuing delocalization of the orientation state is counteracted by orientational decoherence processes \cite{stickler2016b,zhong2016} which potentially suppress the revivals. As in other matter-wave experiments \cite{hornberger2003a,hackermuller2004,romeroisart2011a,bateman2014,pino2018}, the dominant sources of environmental decoherence are the scattering of residual gas atoms and the thermal emission of photons.

Exactly solving the Markovian quantum master equation of orientational decoherence \cite{stickler2016b,zhong2016} is challenging due to the vast number of occupied angular momentum states. As a conservative estimate, we assume that a single decoherence event suffices to completely destroy the alignment signal, by producing a state $\rho_{\rm i}$ with $\matel{\Omega}{\rho_{\rm i}}{\Omega} = 1/4 \pi$ {\color{RoyalBlue}where $\Omega$ is the rotor orientation}. The rotor dynamics can then be described by the master equation $\partial_t \rho = -i[\mathsf{H}_0,\rho]/\hbar+\Gamma\left( \rho_{\rm i} -\rho\right)$, where $\Gamma$ is the total rate of decoherence events. The alignment $\mitl{\cos^2 \beta}=\mathrm{tr}[\rho(t)\cos^2\hat{\beta}]$ at time $t$ follows as 
\begin{equation} \label{eq:cos2deco} 
\mitl{\cos^2 \beta} = \mitl{\cos^2 \beta}_{\rm u} e^{-\Gamma t} + \frac{1}{3} \left ( 1 - e^{-\Gamma t} \right ),
\end{equation}
where $\mitl{\cos^2 \beta}_{\rm u} = \mathrm{tr}[\rho_{\rm u}(t)\cos^2\hat{\beta}]$ denotes the alignment dynamics of the decoherence-free evolution (see \ref{app:D}), and
\begin{eqnarray} \label{eq:unievol}\fl
 \matel{j m}{\rho_{\rm u}(t)}{j'm'} = & \sum_{j,j' = 0}^\infty \sum_{m = -j}^{j}  \sum_{m' = -j'}^{j'} \matel{j m}{\rho_0}{j'm'} \exp \left [ - \frac{i \hbar t}{2 I} \left [ j(j+1) - j'(j'+1) \right ]\right ].
\end{eqnarray}

\begin{figure*}
 \centering
 \includegraphics[width = 0.95\textwidth]{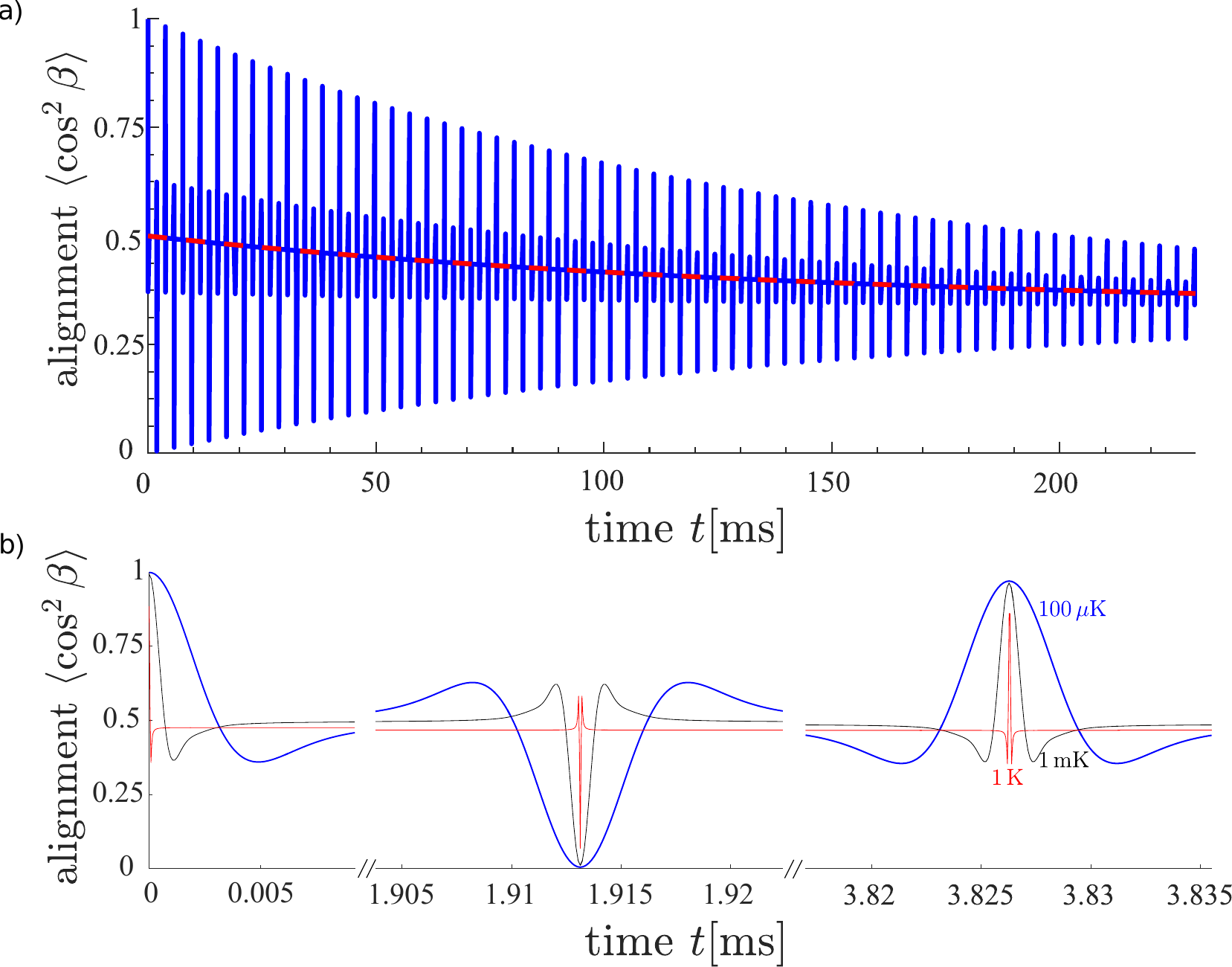}
 \caption{{\color{RoyalBlue}a)} Orientational alignment signal $\mitl{\cos^2\beta}$  of carbon nanotubes ($M=1.9\times 10^5\,$amu, $\ell=50$\,nm) as a function of time (blue solid line) for $T=100\,\mu$K. For most of the time, the alignment shows the classical behavior (dashed red line) of  decaying exponentially with rate $\Gamma$ from 1/2 towards 1/3. However, the initial alignment recurs at integer multiples of the quantum revival time $T_{\rm rev}=2\pi I/\hbar\simeq3.8\,$ms and approaches a minimum at half integer multiples of $T_{\rm rev}$.  {\color{RoyalBlue}b) Initial alignment decay, half-revival and revival for three different initial temperatures.} The width of the revival peaks is determined by the decay time of the initial state, 1/$\kappa= \sqrt{I/2 k_{\rm B} T}$. For {\color{RoyalBlue}$T=100$\,$\mu$K and $T = 1$\,mK} the initially trapped state is numerically calculated by exact diagonalization of $\oH$.} \label{fig:fig2}
\end{figure*}

The initially trapped orientation state quickly disperses during free fall due to its angular momentum spread, {\color{RoyalBlue} see Fig.~\ref{fig:fig2}}. This rapid alignment decay can be approximated using the shearing dynamics associated with a flat orientation space (see \ref{app:C}),
\begin{equation} \label{eq:cos2cl}
 \mitl{\cos^2 \beta}_{{\rm u}} \simeq \mitl{\cos^2 \beta}_0 e^{-\kappa^2 t^2} + \frac{1}{2} \left ( 1 - e^{-\kappa^2 t^2} \right ), 
\end{equation}
with rate $\kappa = \sqrt{2 k_{\rm B} T/I}$.

The corresponding classical dynamics exhibits the same alignment decay since $\rho_0$ is virtually indistinguishable from a classical thermal state for the considered temperatures. After this initial alignment reduction to a value of 1/2, the classically expected alignment shows no revivals at all. Rather, it decays as $1/3  + e^{- \Gamma t}/6$, based on the same assumptions that lead to Eq.~\eref{eq:cos2deco}.

The alignment of the quantized rotor follows this classical prediction for most of the time. However, 
the initial orientation state Eq.~\eref{eq:inalign}  recurs at integer multiples of the revival time $T_{\rm rev} = 2 \pi I / \hbar$, as follows directly from  Eq.~\eref{eq:unievol} and as displayed in Fig.~\ref{fig:fig2}. The width and height of the orientational quantum revival are determined by the initial {\color{RoyalBlue}librational} temperature as described by Eqs. \eref{eq:inalign} and \eref{eq:cos2cl}. {\color{RoyalBlue}In general, quantum revivals can occur when the energy depends quadradically on the quantum number \cite{robinett2004}, such as for the quantum particle in a box or for photons in a Kerr nonlinear medium \cite{yurke1986,sanders1992}}.

\subsection{Measuring the alignment} 

The {\color{RoyalBlue}instantaneous} alignment at variable times can be measured by illuminating the rotor with a {\color{RoyalBlue}weak plane-wave} probe {\color{RoyalBlue}pulse of nanosecond duration} and collecting the scattered light, as demonstrated in Ref.~\cite{kuhn2015}. {\color{RoyalBlue}Using a plane-wave laser ensures that light scattering is independent of the rotor center-of-mass position. Choosing the pulse polarization in the same direction as that of the trapping laser, the total light scattered from the probe laser pulse is proportional to the instantaneous alignment $\langle \cos^2 \beta\rangle$ \cite{kuhn2015,stickler2016a}. The scattering of probe photons decoheres the rotor state to a definite orientation, implementing a projective quantum measurement of $\cos^2 \hat \beta$. By probing the particle orientation after variable times $t$ in repeated experimental runs one can thus test for the emergence of orientational revivals. Note that the optical torque exterted by the probe laser is irrelevant since the rotor is recaptured and re-cycled in the next step.}



\subsection{Recapture}

In the final step of the scheme the rotor is recaptured by switching on the trapping laser when the particle traverses an antinode. Moderate laser powers of a few tens of Watts suffice for a nanoparticle of length $\ell = 50$~nm and mass $M = 10^6$~amu. For such particles the revival time is as short as $T_{\rm rev} \simeq 21$~ms so that the rotor drops only $2.1$~mm and reaches the center-of-mass velocity $0.2$~m/s. For longer revival times, it can be beneficial to use the probe pulse for recapturing the rotor.

Note that the current proposal requires no diffraction grating, the position from which the rotor is released does not affect the alignment signal, and the particle can be recycled during the experiment. These advantages substantially reduce the requirements on nanoparticle fabrication and on source stability as compared to center-of-mass superposition tests.

\section{Discussion}

\subsection{Carbon Nanotubes and Silicon Nanorods} 

To demonstrate the viability of the proposed scheme, we discuss the experimental realization for two types of nanoscale rotors, semiconducting double-walled carbon nanotubes (CNTs) and silicon nanorods  (SNRs). Both can be fabricated with a length of $\ell = 50$~nm, implying that the CNTs have a mass of $M = 1.9\times10^5$~amu (outer diameter $d = 1.5$~nm, inner diameter 1.0\,nm) while the SNRs have a mass of $M =1.4\times 10^6$~amu ($d = 5$~nm). The resulting revival times are $T_{\rm rev} \simeq 3.8$~ms and $T_{\rm rev} \simeq 28$~ms, respectively, implying that the particles fall about $72$~$\mu$m and $4.0$~mm during the first revival. The polarizability anisotropy of CNTs is taken from Ref.~\cite{kozinsky2006}, and that of SNRs is determined as in Refs.~\cite{kuhn2015,stickler2016a}.

The nanoparticle is initially trapped in an optical tweezer of waist $w = 30$~$\mu$m and power $P = 5$~W. For SNRs it has been demonstrated in Ref.~\cite{bateman2014} that internal heating and photon emission is negligible for a wavelength of $1.55$~$\mu$m. For CNTs, where excitonic excitations play no role for wavelengths well above $2.5$~$\mu$m \cite{liu2012}, the exact position and width of the vibrational excitations depends on the structural details of the particle \cite{kim2005} and the optimal trapping wavelength can be determined experimentally.

Cavity or feedback cooling the rotation of the trapped particles to subkelvin temperatures is feasible \cite{stickler2016a,zhong2017}, but may require low mode volume cavities to enhance the nanoparticle-light interaction and detection efficiency. The deeply trapped particle librates harmonically, so that well established techniques of center-of-mass optical cooling can be adapted \cite{gieseler2012,kiesel2013,asenbaum2013,millen2015}.

The total decoherence rate $\Gamma$ accounts for collisions with residual gas atoms and for the emission of thermal photons, $\Gamma = \Gamma_{\rm gas} + \Gamma_{\rm emi}$. The rate at which thermal gas atoms of mass $m_{\rm g}$, pressure $p_{\rm g}$, and temperature $T_{\rm g}$ scatter off a cylinder of length $\ell$ and effective diameter $d_{\rm eff}$ can be estimated by integrating the mean particle flux into the surface over the particle shape
\begin{equation}
\Gamma_{\mathrm{gas}}=\frac{\pi p_\mathrm{g} d_{\rm eff} \ell}{\sqrt{2\pi m_\mathrm{g} k_{\rm B} T_{\rm g}}} \left (1 +  \frac{d_{\rm eff}}{2 \ell} \right).
\end{equation}

The rate of thermally emitted photons depends on the internal temperature and the material-specific spectral absorption cross section of the nanoparticle \cite{hansen1998,bateman2014}. The former is determined by the internal heating of the particle during the recycling and alignment step. The effect of heating can be minimized by choosing the infrared wavelength of the trapping laser between vibrational transitions, where the particle is practically transparent.

Figure~\ref{fig:fig2} shows the expected alignment for CNTs as a function of the time delay between nanoparticle release and detection, as numerically calculated by propagating the initial state. The  latter is determined by exact numerical diagonalization of $\oH$, involving about two hundred thousand total angular momentum quanta. The simulation shows that the alignment approaches a minimum at all half integer multiples of $T_{\rm rev}$, a quantum effect related to the angular momentum parity of the initial state, and decays on the timescale $1/\Gamma\simeq 145\,$ms due to collisional decoherence at $p_{\rm g} = 5\times 10^{-9}$~mbar (assuming $d_{\rm eff}=2d$). The expected signal for SNRs displays a similar structure, as shown in \ref{app:E}. In both cases the orientational quantum revivals are clearly visible, demonstrating that their observation is an achievable goal in the near future.

\subsection{Macroscopicity} 

In order to assess the proposed superposition test we consider the empirical measure of macroscopicity $\mu$ as defined in Ref.~\cite{nimmrichter2013}. It quantifies to what extent a successful superposition experiment serves to rule out a wide class of classicalizing modifications of quantum theory. The resulting dynamics can be solved exactly for planar rotations (see \ref{app:F}), which provides a lower bound for the macroscopicity of the $n$-th linear rotor quantum revival,
\begin{equation}
\mu \geq \log_{10} \left [\frac{n \theta_{\rm m}}{\vert\ln f \vert} \left ( \frac{M}{m_e} \right )^2 \frac{T_{\rm rev}}{1~{\rm s}} \right ],
\end{equation}
where $f$ is the ratio of observed-to-expected signal visibility and $\theta_m \simeq 0.12$ is a numerical factor. Assuming $M = 10^6$~amu, $\ell = 50$~nm, and $f = 0.8$ at the tenth revival yields a lower bound of $\mu \geq 17.5$, on a par with ambitious center-of-mass interference proposals.  For comparison, a tobacco mosaic virus \cite{bruckman2014} with $\ell \simeq 300$~nm, $d \simeq 20$~nm, and $M \simeq 4\times 10^7$~amu has $T_{\rm rev} \simeq 30$~s, and observation of its first revival ($f = 0.8$) would imply $\mu \geq 22.8$.

\begin{figure*}
 \centering
 \includegraphics[width = 0.7\textwidth]{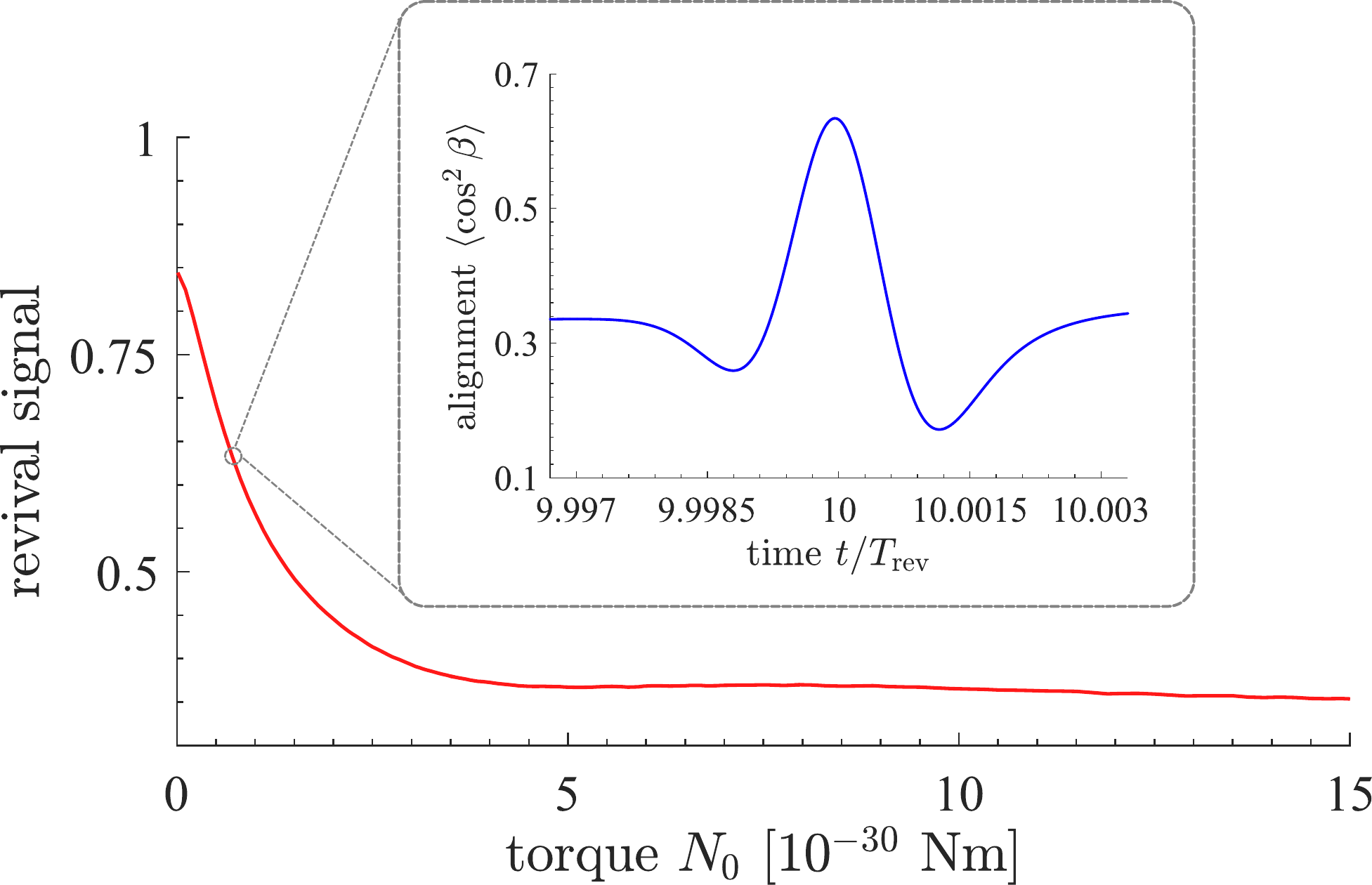}
 \caption{The decay of the tenth revival signal of a CNT ($T = 100~\mu$K) in presence of an external torque $N_{\rm ext}$ can be used for ultra-precise sensing, surpassing the sensitivity of state-of-the-art systems by many orders of magnitude. The inset shows the alignment signal as a function of time for $N_{\rm ext} = 7\times 10^{-31}$~Nm.} \label{fig:fig3}
\end{figure*}

\subsection{Sensing Applications} 

The presence of an external torque during free fall can have a strong  influence on the orientational revival signal. By monitoring the alignment as a function of time for different initial orientations the magnitude and direction of an applied torque can thus be deduced. The torque sensitivity of the tenth revival of a CNT ($T = 100~\mu$K) is illustrated in Fig.~\ref{fig:fig3}, showing the alignment reduction due to an external torque of magnitude $N_{\rm ext}$ orthogonal to the trapping laser polarization. The numerical simulations of the torque-induced dynamics, involving transitions between 360,000 angular momentum states (see Materials and Methods), show that torques on the order of $10^{-30}$~Nm are observable, eight orders of magnitude smaller than the levitated \cite{hoang2016,kuhn2017b} or solid-state-integrated \cite{wu2017} setups considered so far.

By attaching single elementary charges to the ends of the silicon nanorod discussed above one can measure electrostatic fields at values well below mV/m. The decay of the alignment signal when the rotor is exposed to an atomic beam or other controlled environments can be used for studying collisional decoherence and thermalization of quantum nanoscale rotors. Finally, objective collapse models could be tested by observing orientational quantum revivals that contradict the predicted loss of orientational coherence \cite{schrinski2017}.

\section{Conclusion}

We presented a viable scheme for the first observation of orientational quantum revivals of nanoscale particles. The proposed experiment can be realized with upcoming technology, representing a macroscopic test of the superposition principle and opening the door to quantum enhanced torque sensing. The successful demonstration of orientational revivals may well be the starting point for interferometric manipulation methods of nanoscale rigid rotors, based on applying a sequence of optical potentials during the free evolution.

\ack
B.A.S. and B.P. contributed equally.
This work was supported by the Deutsche Forschungsgemeinschaft (DFG -- 394398290) and by the Austrian Science Fund
(FWF, P27297). S.K.\ acknowledges funding from the ESQ Discovery Grant ROTOQUOP of the
Austrian Academy of Sciences ({\"O}AW).

\appendix
\section{Semiclassical evaluation of matrix elements}\label{app:A}

For completeness, we briefly summarize how to approximate quantum mechanical matrix elements of the linear rotor using Bohr-Sommerfeld quantization. The angles $\alpha, \beta$ and their canonical angular momenta $p_\alpha,p_\beta$  are related to the action-angle variables $\alpha_m,\alpha_j,m,j$ via \cite{child2014}
	\begin{eqnarray}\label{eq:replace}
	\alpha&=& \alpha_m +  \mathrm{arctan}\left( \zeta \tan \alpha_j \right ) - \pi
\\
	\cos \beta &=& \cos \alpha_j \sqrt{1 - \zeta^2}
\\	\sin \beta &=& \sqrt{ \sin^2 \alpha_j + \zeta^2 \cos^2 \alpha_j}
\\
	p_\beta &= & \hbar \left (j + \frac{1}{2} \right ) \frac{\sin \alpha_j \sqrt{1 - \zeta^2}}{\sqrt{ \sin^2 \alpha_j + \zeta^2 \cos^2 \alpha_j}},
\\
	p_\alpha &=& \hbar m,
	\end{eqnarray}
with $\zeta = m/(j + 1/2)$. Note that these relations imply $p_\alpha^2/\sin^2 \beta + p_\beta^2 = \hbar^2 (j+1/2)^2$.

The matrix elements $\matel{jm}{\op{A}}{j'm'}$ of an arbitrary operator $\op{A} = A(\hat{\alpha}, \hat{\beta}, \op{p}_\alpha, \op{p}_\beta)$ can be semiclassically approximated by first replacing all arguments of $A$ according to Eq.~\eref{eq:replace}, yielding $\overline{A}(\alpha_m,\alpha_j,m,j)$. The matrix elements can then be obtained by calculating
	\begin{eqnarray}\fl
	\matel{j m}{\op{A}}{j'm'} \simeq \frac{1}{(2 \pi)^2} \int_0^{2\pi} d\alpha_m \int_0^{2 \pi}d\alpha_j e^{i \alpha_m (m - m')} e^{i \alpha_j ( j - j')} \overline{A} \left ( \alpha_m,\alpha_j ,\frac{m+m'}{2},\frac{j + j'}{2} \right ).
	\nonumber\\
	\end{eqnarray}
Applying this to $\rho = \exp( - \oH/k_{\rm B} T)/Z$ yields Eq.~\eref{eq:scis}.

\section{ Initial Alignment}\label{app:B}

In order to estimate the initial alignment of a rotor in the potential $V(\beta) = - V_0 \cos^2\beta$ we calculate the expectation value $\mitl{\cos^2\beta}_0 = \mathrm{tr}(\rho_0 \cos^2 \hat{\beta})$. Inserting the initial state $\rho_0 = \exp ( - \oH / k_{\rm B }T) /Z$ with $\oH = \oJ^2/2 I + V(\hat{\beta})$ shows that the expectation value can be expressed in terms of the partition function $Z$,
\begin{equation} \label{eq:zrel}
\mitl{\cos^2 \beta}_0 = k_{\rm B} T \frac{\partial}{\partial V_0} \ln Z.
\end{equation}
The latter can be calculated explicitly in the semiclassical limit, where the matrix elements of $\rho_0$ take the form Eq.~\eref{eq:scis},
\begin{eqnarray}
	Z = &\sum_{j = 0}^\infty \sum_{m = -j}^j \exp \left [ - \frac{\hbar^2 (j + 1/2)^2}{2 I k_{\rm B} T} \right ] I_0 \left [ \frac{V_0}{2 k_{\rm B} T} \left ( 1 - \frac{m^2}{(j + 1/2)^2} \right ) \right ] \nonumber\\
	&\times\exp \left [ \frac{V_0}{2 k_{\rm B} T} \left ( 1 - \frac{m^2}{(j + 1/2)^2} \right ) \right ].
\end{eqnarray}

One then replaces the sum over $m$ by an integral from $-(j+1/2)$ to $+(j+1/2)$ and the sum over $j$ by an integral over $j+1/2$. After the substitution $u = m/(j+1/2)$ we thus obtain
\begin{equation}
Z \simeq \frac{I k_{\rm B} T}{\hbar^2} \int_{-1}^1 du\, I_0 \left [ \frac{V_0}{2 k_{\rm B} T} (1 - u^2) \right ] \exp  \left [ \frac{V_0}{2 k_{\rm B} T} (1 - u^2) \right ].
\end{equation}

The integral can be evaluated for $k_{\rm B} T/V_0 \ll 1$ by using the asymptotic expansion $I_0(z) \sim e^z/\sqrt{2 \pi z}$ as $z \to \infty$, 
\begin{equation}
Z \simeq \frac{I (k_{\rm B} T)^2}{\hbar^2V_0}\exp \left ( \frac{V_0}{k_{\rm B} T} \right ).
\end{equation}
Inserting this into Eq.~\eref{eq:zrel} gives Eq.~\eref{eq:inalign}.

We remark that Eq.~\eref{eq:inalign} can also be obtained classically by calculating $\mitl{\cos^2 \beta}_0$ with the marginal Boltzmann distribution of the polar angle $\beta$ in the trap $f(\beta) = \sin \beta \exp ( V_0 \cos^2 \beta / k_{\rm B} T) / Z'$.

The total angular momentum expectation value of the initial distribution can be estimated with the semiclassical substitutions used above. In particular, carrying out the integral over $m$ and subsuming the result into the new normalization $Z'$ yields
\begin{equation}
\mitl{j}_0 \simeq \frac{1}{Z'} \int_0^\infty dj\, j^2 \exp \left ( - \frac{\hbar^2 j^2}{2 I k_{\rm B} T} \right ) = \sqrt{\frac{\pi I k_{\rm B} T}{2 \hbar^2}}.
\end{equation}

\section{Decoherence free alignment dynamics}\label{app:D}
The time-dependent alignment $\mitl{\cos^2 \beta}_{\rm u}$ due to the unitary dynamics is numerically calculated by carrying out the trace over the product of the time evolved state Eq.~\eref{eq:unievol} and the operator-valued observable $\cos^2 \hat{\beta}$,
	\begin{equation}
	\left\langle\cos^2\beta\right\rangle_{\rm u}= \sum_{j = 0}^\infty \sum_{m = -j}^j \sum_{j' = 0}^\infty \sum_{m' = -j'}^{j'} \matel{j m}{\rho_{\rm u}(t)}{j'm'}\matel{j'm'}{\cos^2 \hat{\beta}}{jm},
	\end{equation}
	where the matrix elements of $\rho_{\rm u}(t)$ are given by Eq.~\eref{eq:unievol} and
	\begin{eqnarray} \label{eq:cos2qu}\fl
	\left\langle jm\right\vert\cos^2\hat{\beta}\left\vert j'm'\right\rangle=& \frac{\delta_{mm'}}{3} \Bigg[ (-1)^m2\sqrt{(2j'+1)(2j+1)}\left(\begin{array}{c c c} j' &2&j\\
	0&0&0\end{array}\right)\left(\begin{array}{c c c} j' &2&j\\
	m&0&-m\end{array}\right)
	\nonumber\\
	& +\delta_{jj'} \Bigg ]
	\end{eqnarray}
Here, the angular brackets denote Wigner-3j symbols \cite{brink2002}. Due to their selection rules, Eq.~\eref{eq:cos2qu} vanishes unless $j = j', j' \pm 2$, which significantly simplifies the evaluation of the alignment \eref{eq:cos2deco}.

\section{ Classical Dispersion}\label{app:C}
We estimate the dispersion timescale by calculating the classical alignment loss of a Gaussian state released from the laser potential $V_0$ and approximating the rotor dynamics as flat. On a short timescale, the angle $\beta$ evolves to $\beta(t) \simeq \beta + p_\beta t/I$, where $p_\beta$ is the corresponding angular momentum. The marginal distribution of $\beta$ and $p_\beta$ follows from the Boltzmann distribution as $g(\beta,p_\beta) = \sin \beta \exp ( - p_\beta^2/2 I k_{\rm B} T + V_0 \cos^2 \beta/ k_{\rm B} T )/N$. With this one obtains the expectation value
\begin{eqnarray}
\mitl{\cos^2 \beta}_{\rm u} & \simeq  \int_0^\pi d\beta \int_{-\infty}^\infty dp_\beta\, g(\beta,p_\beta) \cos^2 \left ( \beta + \frac{p_\beta t}{I} \right ) \nonumber \\
&= \mitl{\cos^2 \beta}_0 e^{-\kappa^2 t^2} + \frac{1}{2} \left ( 1 - e^{-\kappa^2 t^2} \right ),
\end{eqnarray}
with $\kappa = \sqrt{2 k_{\rm B} T/I}$.

\section{ Orientational revivals of silicon nanorods}\label{app:E}

\begin{figure*}
	\centering
	\includegraphics[width = 0.7\textwidth]{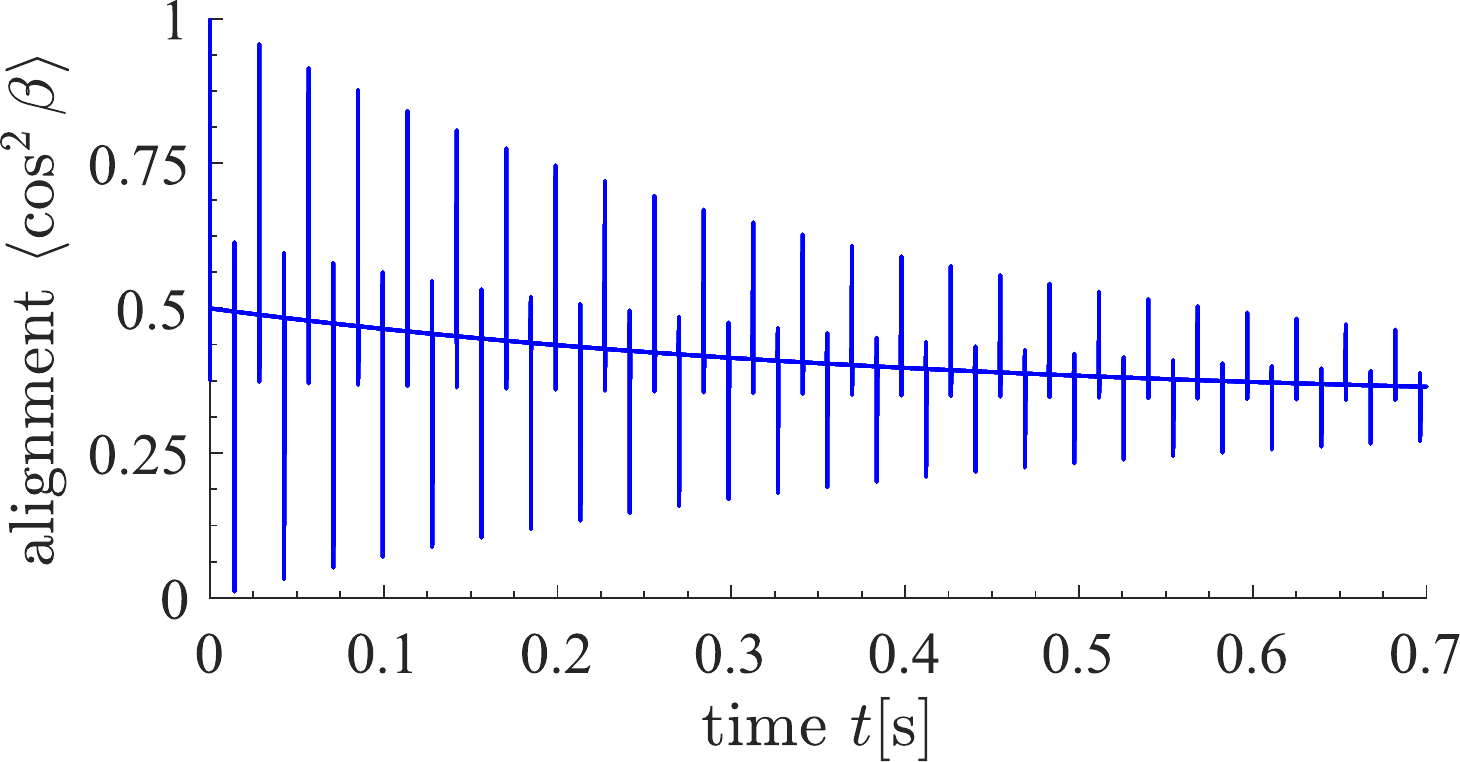}
	\caption{Orientational alignment $\mitl{\cos^2\beta}$ of the silicon nanorod discussed in the main text, as a function of time at $T=100\,\mu$K. For most of the times, the alignment signal decays exponentially with decoherence rate $\Gamma$ from 1/2 towards 1/3. However, the initial alignment recurs at integer multiples of the revival time $T_{\rm rev}=2\pi I/\hbar\simeq28$\,ms, and approaches a minimum at half integer multiples of $T_{\rm rev}$.  } \label{fig:supfig}
\end{figure*}

Figure \ref{fig:supfig} shows the alignment signal for silicon nanorods as discussed in the main text, at a gas pressure of $p_{\rm g}=5\times 10^{-10}$\,mbar. As in the case of carbon nanotubes, many revivals can be observed. The revival time is $T_{\rm rev}=28\,$ms.

\section{ Macroscopicity}\label{app:F}
In order to estimate the macroscopicity of the superposition state reached in the proposed experiment one replaces the unitary time evolution of the rotation state $\rho$ between particle release and revival by the dynamics described by the quantum Markovian master equation  $\partial_t \rho = -i\left [ \oH,\rho \right ]/\hbar + \cL \rho$ discussed in \cite{nimmrichter2013}. In the present case, the  superoperator $\cL \rho$ takes the form
\begin{eqnarray} \label{eq:mod}
\cL \rho = &\frac{1}{\tau m_e^2} \frac{1}{ \left ({2 \pi \sigma_q^2} \right )^{3/2}} \int d^3 {\bf q} \exp \left  ( - \frac{q^2}{2 \sigma_q^2} \right ) \Bigg( \tilde{\varrho}\left [ \rR^T(\oOmega) {\bf q} \right ] \rho \tilde{\varrho}^*\left [ \rR^T(\oOmega) {\bf q} \right ]
\nonumber\\
 &- \frac{1}{2} \left \{ \left \vert \tilde{\varrho}\left [ \rR^T(\oOmega) {\bf q} \right ] \right  \vert^2, \rho \right \} \Bigg),
\end{eqnarray}
since it is permissible to neglect position displacements and to approximate the particle by a homogeneous mass density. Here, $\tau$ gives the time scale on which the modification acts, $m_e$ is the electron mass, $\sigma_q$ is the width of the momentum-kick distribution, $\rR(\oOmega)$ is the operator-valued rotation matrix, and $\tilde{\varrho}\left ({\bf q} \right )$ is the Fourier transform of the mass density. This master equation is similar to that for rotational collapse dynamics in the model of continuous spontaneous localization \cite{schrinski2017}. For a homogeneous rod of mass $M$, length $\ell$, and whose symmetry axis points into direction ${\bf m}(\Omega) = \rR(\Omega) {\bf e}_z$, one obtains
\begin{equation}
\tilde{\varrho}\left [ \rR^T(\Omega) {\bf q} \right ] \simeq M {\rm sinc} \left [ \frac{\ell}{2\hbar} {\bf q} \cdot {\bf m}(\Omega) \right ].
\end{equation}

The master equation can be  solved exactly if the rotor is confined to a plane so that its orientation is characterized by a single angle $\alpha$ and ${\bf m}(\Omega) = {\bf e}_\rho(\alpha)$. The resulting macroscopicity of the planar rotor is then a lower bound to that of the linear top. The solution of the master equation can be calculated by using the discrete phase space of the orientation state \cite{schrinski2017}. If the rotor is initially perfectly aligned, $\langle \cos^2 \alpha \rangle_0 = 1$, the alignment at an integer multiple of the linear rotor revival time, $t = n T_{\rm rev}$, is reduced by the modificiation Eq.~\eref{eq:mod} according to
\begin{equation} \label{eq:macro}
\langle \cos^2 \alpha \rangle_{n} = \frac{1}{2} + \frac{1}{2} \exp \left [ - n \frac{T_{\rm rev}}{\tau} \theta \left ( \frac{\ell \sigma_{q}}{\hbar} \right )  \left ( \frac{M}{m_e} \right )^2 \right ],
\end{equation}
where
\begin{eqnarray} \label{eq:theta}
\theta(x) =& \frac{1}{2} \int_0^\infty d u~ u e^{-u^2/2} \int_0^{2 \pi} \frac{d \varphi}{2 \pi} \int_0^{2 \pi}\frac{d \alpha}{2 \pi}
\nonumber\\ 
&\times\left [ {\rm sinc}\left [ \frac{u x}{2} \cos \left ( \alpha - \frac{\varphi}{2} \right ) \right ] - {\rm sinc}\left [ \frac{u x}{2} \cos \left ( \alpha + \frac{\varphi}{2} \right ) \right ]\right ]^2.
\end{eqnarray}

The macroscopicity associated with the measured alignment signal $\langle \cos^2 \alpha \rangle_n = (1 + f)/2$ is determined by inserting the maximum of Eq.~\eref{eq:theta}, $\theta_{\rm m} \simeq 0.12$, into Eq.~\eref{eq:macro}, solving for $\tau$ and taking the decadic logarithm. Here  $f = 2 \langle \cos^2 \alpha \rangle_n - 1$ is the ratio of observed to expected signal visibility.

\section{Torque Sensing}\label{app:G}

In order to quantitatively assess the sensing potential of the proposed revival experiment, the free time evolution \eref{eq:unievol} has to be replaced by that in presence of an external torque of magnitude $N_{\rm ext}$. For illustration, we consider a torque acting orthogonal to the polarization direction of the trapping laser, so that the  operator-valued potential reads $\oV_{\rm ext} = -N_{\rm ext} \sin^2 \hat{\beta} \cos^2 \hat{\alpha}$ with the azimuth $\alpha$.

The exact time evolution can in principle be obtained by numerically diagonalizing the rotational Hamiltonian $\oH = \oJ^2/2I + \oV_{\rm ext}$ to describe transitions between all relevant rotation states, which amounts to approximately 400,000 states for a $T = 100 ~\mu$K CNT. Since {\it all} resulting eigenvalues and eigenvectors must be stored, this method soon becomes numerically expensive, scaling as $J^4$ with the maximal angular momentum quantum number $J$. However, in the present case one can use that the considered torques $N_{\rm ext}$ are much smaller than the kinetic energy of the occupied states to apply degenerate perturbation theory.

The eigenergies of the Hamiltonian $\oH$  can be obtained to first order in $N_{\rm ext}$ by diagonalizing for each $j$ the matrices
\begin{equation}
 {\rm H}_j = \frac{\hbar^2}{2 I} j(j+1) \un_j + {\rm V}_j,
\end{equation}
where $\un_j$ is a $(2j+1) \times (2j+1)$ identity matrix and $({\rm V}_j)_{mm'} = \matel{jm}{\oV_{\rm ext}}{jm'}$. The resulting eigenenergies $E_{jn}$ are generally non-degenerate and labeld by $n = -j,\ldots, j$  for each $j$. The eigenergies are thus no longer an integer multiple of $2 \pi \hbar/T_{\rm rev}$, leading to the accumulation of a phase, which causes the revival to diminish. Our simulations show that this phase accumulation is the dominant contribution to the revival decay, while the perturbation of the eigenstates can be ignored.

We thus replace the exact unitary time evolution by the $j$-conserving evolution operator ${\rm U}_j = \exp \left ( - i t {\rm H}_j/\hbar \right )$, acting on the subspace of fixed $j$, thereby reducing the above scaling to $J^3$. Diagonalization of the initial state, $\rho_0 = \sum_k p_k \ketbra{\psi_k}{\psi_k}$, then yields the alignment signal in the numerically tractable (but still expensive) form
\begin{equation}
\mitl{\cos^2 \beta} = \sum_k p_k \sum_{j,j' = 0}^\infty \underline{\Psi}_{kj}^* \cdot {\rm U}_j {\rm C}_{jj'} {\rm U}^\dagger_{j'} \underline{\Psi}_{kj'},
\end{equation}
where ${\rm C}_{jj'}$ is the $(2j+1) \times (2j'+1)$-matrix with elements Eq.~\eref{eq:cos2qu} and we defined the vector $(\underline{\Psi}_{kj})_m = \braket{jm}{\psi_k}$. This expression can be simplified further by exploiting that ${\rm C}_{jj'}$ is only non-zero for $j = j'$ and $j = j' \pm 2$.


\end{document}